\documentclass[fleqn,twoside]{article}
\usepackage{espcrc2}

\usepackage{amsmath}
\usepackage{amssymb}
\usepackage{graphicx}

\newcommand{\av}[1]{\langle #1 \rangle}
\newcommand{\xv}{{\bf x}}

\newcommand{\ver}{{\bf \hat n}}

\newcommand{\tr}{\operatorname{Tr}}

\newcommand{\re}{\operatorname{Re}}

\newcommand{\plx}{v(\xv)}
\newcommand{\pl}{\mathcal{L}}

\title{Dimensional reduction and a Z(3) symmetric model}
\author{
P.~Bialas, \address{Inst. of Physics, Jagellonian University, 33-059 Krakow, Poland}
A.~Morel, \address{Service de Physique Th\'eorique de Saclay, CE-Saclay,
F-91191 Gif-sur-Yvette Cedex, France} 
B.~Petersson, 
\address[BI]{Fakult\"at f\"ur Physik, Universit\"at Bielefeld, P.O.Box 100131, D-33501 Bielefeld, Germany} 
\thanks{The contribution was presented by B.P.} 
}
\begin{document}
\begin{abstract}
We present first results from a numerical investigation of a
$Z_3$ symmetric model based on dimensional reduction.
\end{abstract}

\maketitle

\section{The Model}

In this note we will describe some of the properties of a $Z_3$
symmetric generalization of dimensional reduction, applied to $SU(3)$
gauge theory in $2+1$ dimension.  The motivation is to construct an
effective action, which is constrained by perturbative dimensional
reduction at high temperature, but which can be applied all the way
down to the confinement phase transition. The partition function is
defined as follows:
\begin{equation}\begin{split}\label{eq:ahm-action}
Z&=\int\text{D}[U]\text{D}[V]\exp\bigl(-S_{eff}\bigr)\\
S_{eff}&=S_U+S_{U,V}+S_V
\end{split}
\end{equation}
\begin{align}
  S_U &=\beta_3 L_0 \sum_\Box \bigl(1-\frac{1}{3}\re\tr
  U_\Box\bigr)\\
\label{eq:z3-action-kinetic}
S_{U,V}&=\frac{\beta_{3}}{L_0}\sum_{\xv,i} \bigl(1-\frac{1}{3}
\re \tr U(\xv,i)V(\xv+\ver_i)\notag\\
&\kern20mm U(\xv,i)^\dagger V(\xv)^\dagger\bigr)\\
S_V&=\lambda_2 \sum_{\xv}|\tr V(\xv)/3|^2 \,+\notag\\
&\phantom{=}\kern2mm\lambda_3\sum_{\xv}\re (\tr
V(\xv)/3)^3
\end{align}
The $SU(3)$ matrices $V(\xv)$ and $U(\xv,i)$ are defined on the
sites and the links respectively of a two dimensional lattice. The matrices
$V$ represent the Polyakov loops of the full $(2+1)D$ theory. Similar
models have been proposed in \cite{Pisarski:2000eq}.
Here we discuss the phase structure of the above model, and
compare with results obtained within the standard reduction recipes 
\cite{ahm,Bialas:2000zj}.
For simplicity we set $\lambda_3 = 0$ and discuss the phase
structure in the $(\lambda_2,\beta_3)$ plane. The number $L_0$
of time slices in (2+1)D, which enters as a parameter in $Z$ is fixed to 4
throughout this paper.

\section{Numerical simulations}

The model \eqref{eq:ahm-action} was studied using conventional lattice
QCD Monte-Carlo techniques. The multi hit metropolis algorithm was
used both for updating the gauge fields $U$ and Polyakov loops fields
$V$.  Lattices of sizes ranging from $16\times 16$ to $72\times 72$
were studied.  
The main quantities measured were the traces of  Polyakov loops 
\begin{equation}
\pl=\sum_\xv \plx,\qquad \plx=\frac{1}{3}\tr V(\xv)
\end{equation}
and their two--point correlators.

\section{Results}

\begin{figure}
\begin{center}
\includegraphics[width=6cm]{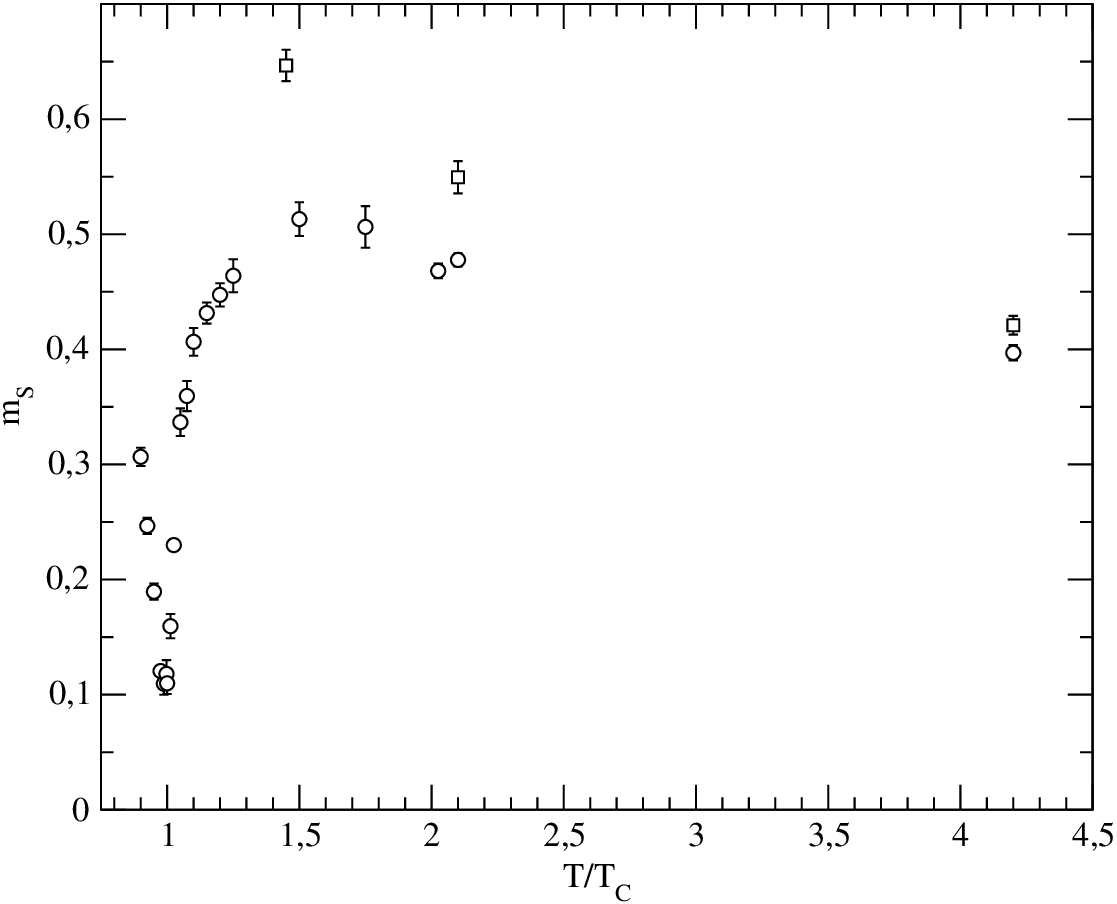}
\end{center}
\caption{\label{fig:massnp}
  Screening masses:  In the present model for
 $\lambda_2=\lambda_3=0$ (circles) and in the naively reduced model (squares,
 see \cite{ahm})
.}
\end{figure}

We started with simulations at $\lambda_2 =0$ for a
range of $\beta_3$ values and got the expected signal of the phase
transition between the low-temperature confined $Z(3)$ symmetric phase
(Ia) and a broken high temperature phase (II).
Screening masses $m_S$ were obtained by fitting the two point Polyakov loop
unconnected correlators  with the formula 
\begin{equation}
P(r)\approx 
a\biggl(\frac{e^{-m_S r}}{\sqrt{r}}+\frac{e^{-m_S (L-r)}}{\sqrt{L-r}}\biggr)+c
\end{equation}
We plot the result of the fit in 
figure~\ref{fig:massnp}. On the same plot we show the results obtained
in \cite{ahm} with the naively reduced model (no Higgs potential).
At sufficiently large temperature (large $\beta_3$) the agreement is good, but 
the $Z_3$ symmetric model also describes the vanishing of the screening
mass associated with $Z_3$ restoration at lower temperatures.

\begin{figure}
\begin{center}
\includegraphics[width=6cm]{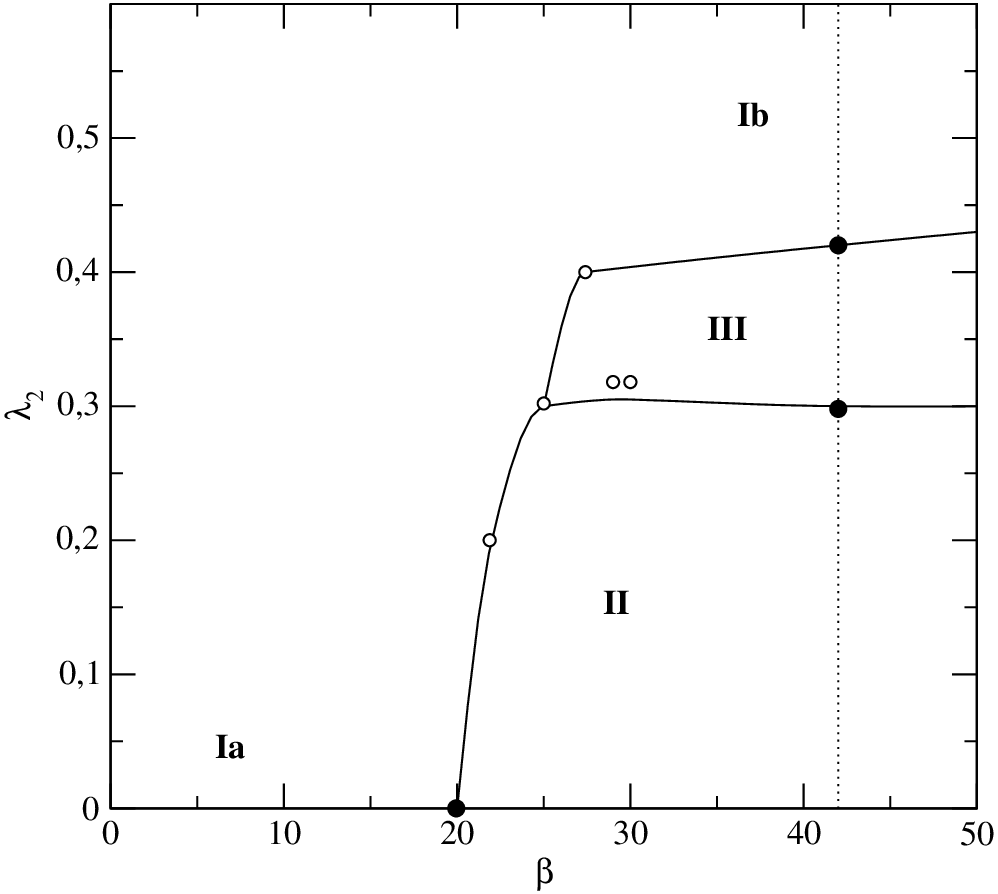}
\end{center}
\caption{\label{fig:pd}
  A tentative phase diagram.}
\end{figure}

\begin{figure}
\begin{center}
\includegraphics[width=6cm]{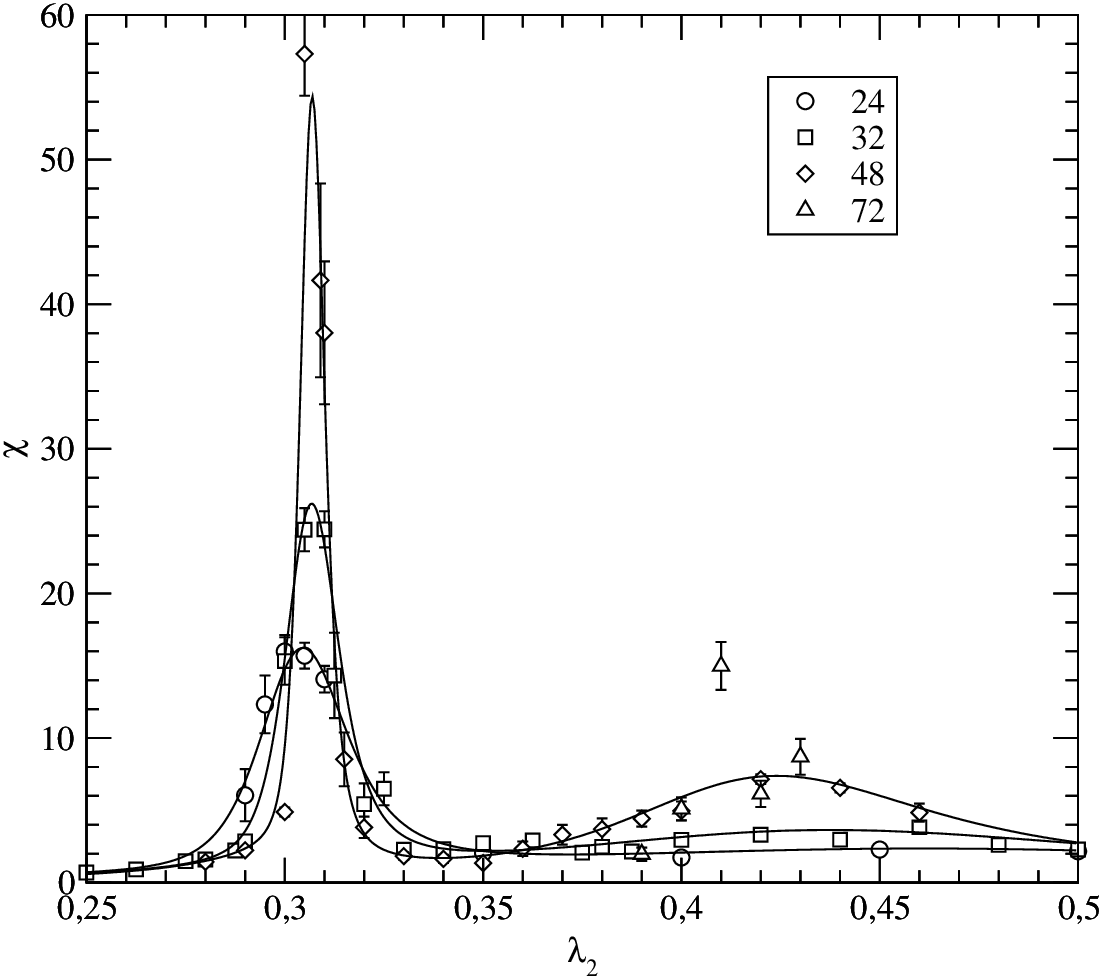}
\end{center}
\caption{\label{fig:xi}Polyakov loop susceptibility as a function of $\lambda_2$ for $\beta_3=42$  on  lattices of various sizes.}
\end{figure}

We then performed a series of scans in the $\beta_3$--$\lambda_2$
plane on $16\times 16$ lattices and looked for peaks in the
susceptibility 
\begin{equation}\label{eq:xi}
\chi=\av{|\pl|^2}-\av{|\pl|}^2
\end{equation}
as indications of phase transitions, while the associated
distributions of $\pl$ were used to characterize the phases. After so
identifying a tentative phase diagram (see figure~\ref{fig:pd}), 
we made a more precise study on larger lattices along the line
$\beta_3=42$.  The results for the susceptibility are
plotted in figure~\ref{fig:xi}.
One sees a strong transition signal just above
$\lambda_2=0.3$. There the size dependence of $\chi$ at its maximum,
as well as the time history of $|\pl|^2$ and its doubly peaked histogram 
clearly indicate that the transition is first order.

\begin{figure}
\begin{center}
\includegraphics[width=5cm]{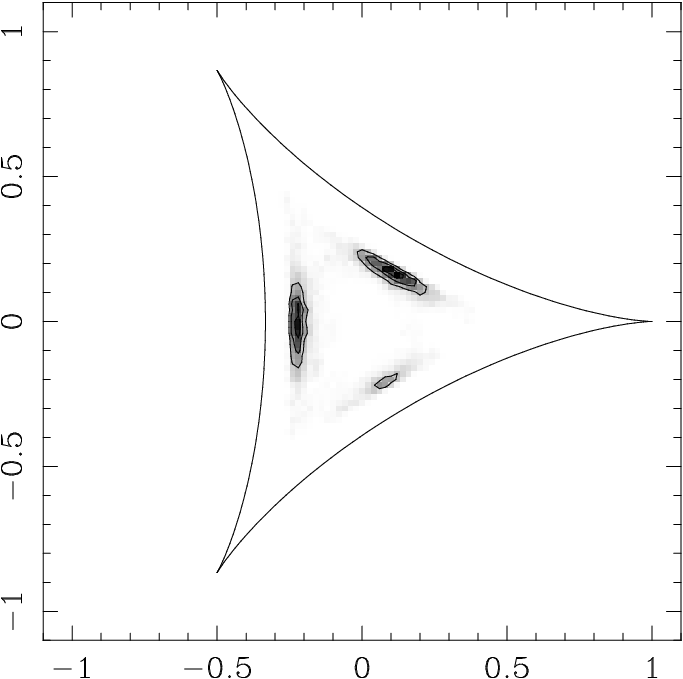}
\end{center}
\caption{\label{fig:hist} Typical distribution of $P(\pl)$ in the
complex plane of $\pl /L^2$ for region III}
\end{figure}

The evidence for a second phase transition, slightly above $\lambda_2
=0.4$ is considerably weaker.
The peak is less pronounced and we had to go to
$72\times 72$ lattices (with much poorer statistics) to make sure 
that it  grows with the lattice size.
In order to study the nature of the phases, we looked at the distributions of
the Polyakov loops $\pl$.  The regions marked Ia and II on 
figure~\ref{fig:pd} are
easily identified with standard $Z(3)$ symmetric and broken phases, with
data concentrated respectively around $\pl \approx 0$ and
$\pl\approx \,\exp(2\,i\,\pi\,n/3)$.
At larger $\lambda_2$ in region III, a new pattern shows up, with again
$Z_3$ breaking peaks, but the argument of $\pl$ is now close to
$(2\,n+1)\,\pi\,/3$,
as illustrated on figure~\ref{fig:hist}. 
Finally, when $\lambda_2$ is further increased (region Ib), the 
histogram is located around $\pl=0$, no particular argument is selected,
and we conjecture that the phase 
is the same as the low temperature confined one (Ia).
Expanding the $U$-fields to second order in the SU(3) algebra and
integrating over the latter support the phase structure just outlined. 
This will be described elsewhere.

\begin{figure}
\begin{center}
\includegraphics[width=6.0cm]{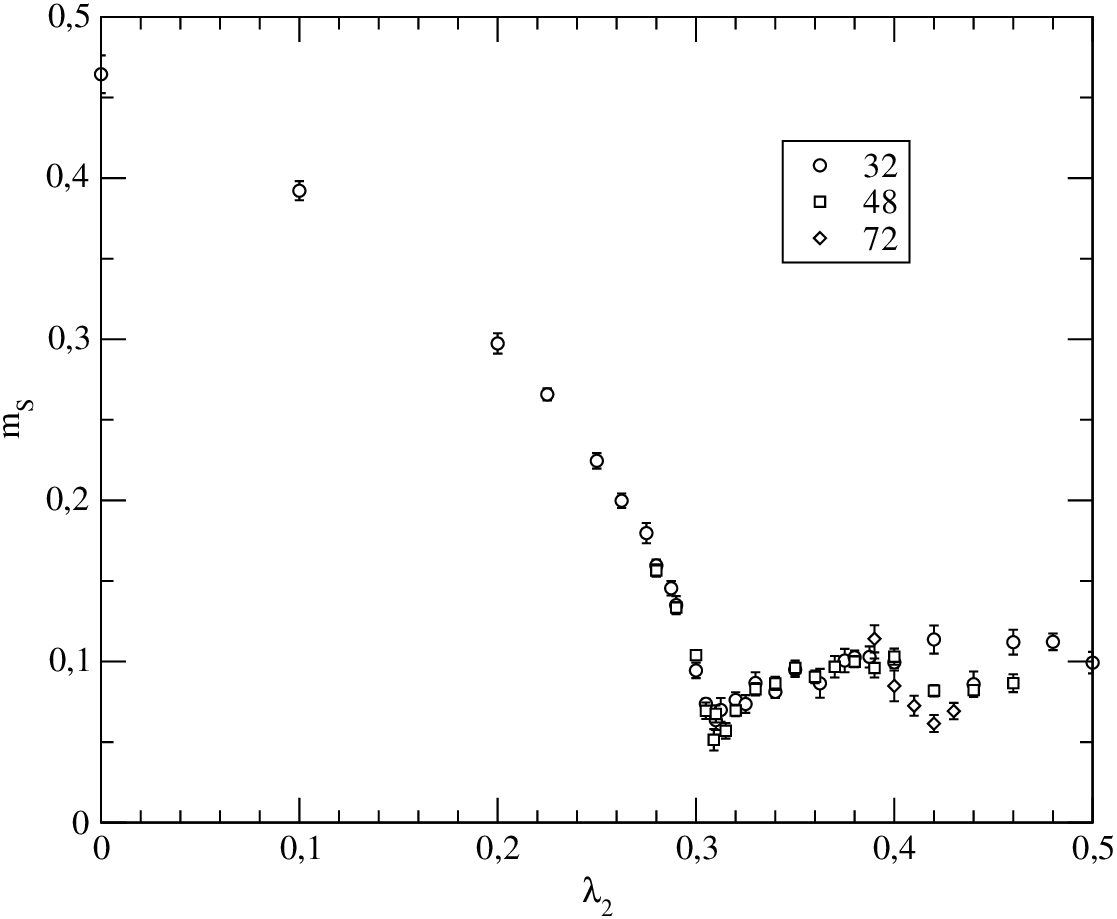}
\end{center}
\caption{\label{fig:mass}
The screening masses $m_S$ versus $\lambda_2$ for $\beta_3=42$ and 
various lattice sizes.}
\end{figure}

Along $\beta_3=42$, 
 we have used the projected (0--momentum) correlators
 to extract the masses, by fitting them to the formula~:
\begin{equation}
P_{prj}(r)\approx 
a\bigl(e^{-m_S r}+e^{-m_S (L-r)}\bigr)+c
\end{equation}
Our results are summarized in figure~\ref{fig:mass}.  
The mass value is found to coincide with that measured at $\beta_3=42$, 
$L_0=4$ in the original(2+1) model \cite{ahm} for $\lambda_2$ close to 0.2.
This point in parameter space is a stable point of the phase diagram, at
variance with what happens in the conventionally reduced model, where
the physical situation corresponds to a metastable point \cite{ahm}.

In conclusion, the $Z_3$ symmetric model opens a possibility to extend
dimensional reduction down to the temperature where confinement is
restored. It exhibits an interesting phase diagram, with a new phase
whose connection to QCD at finite temperature and/or chemical potential 
should be investigated. For this purpose, a study to constrain the
model parameters is under way.


\begin{thebibliography}{99}

\bibitem{Pisarski:2000eq}
R.~D.~Pisarski,
Phys.\ Rev.\ D {\bf 62}, 111501 (2000)
[arXiv:hep-ph/0006205].
\bibitem{ahm}
P.~Bialas, A.~Morel, B.~Petersson, K.~Petrov and T.~Reisz,
Nucl.\ Phys.\ B {\bf 581} (2000) 477
[arXiv:hep-lat/0003004].
\bibitem{Bialas:2000zj}
P.~Bialas, A.~Morel, B.~Petersson, K.~Petrov and T.~Reisz,
Nucl.\ Phys.\ B {\bf 603}, 369 (2001)
[arXiv:hep-lat/0012019].





\end{thebibliography}
\end{document}